\documentclass[a4paper]{jpconf}
\pdfoutput=1

\usepackage{graphicx}
\usepackage[utf8]{inputenc}
\usepackage{color}
\usepackage{braket}
\usepackage{amsfonts}
\usepackage{amssymb}
\usepackage{amsmath}
\usepackage{hyperref}
\usepackage{cleveref}
\usepackage{listings}
\usepackage{xspace}
\usepackage[usenames,dvipsnames]{xcolor}
\usepackage{siunitx}  
\hypersetup{pdfauthor={Marvin Gerlach},pdftitle={tapir ACAT 21 Proceedings}}

\newcommand{\file}[1]{{\color{green!50!blue} \texttt{#1}}}

\definecolor{codegreen}{rgb}{0,0.6,0}
\definecolor{codegray}{rgb}{0.5,0.5,0.5}
\definecolor{codepurple}{rgb}{0.58,0,0.82}
\definecolor{backcolour}{rgb}{0.95,0.92,0.90}

\lstdefinelanguage{tapirconf}
{
  identifierstyle=\color{gray},
  morecomment=[f][\color{black}]{*\ tapir}
}

\lstdefinelanguage{topsel}
{
    morecomment=[s][\color{codegreen}]{(}{)},
    literate={\{}{{\CodeSymbol{\{}}}1
            {\}}{{\CodeSymbol{\}}}}1
            {;}{{\CodeSymbol{;}}}1
}

\lstdefinelanguage{shell}
{
  morekeywords={
    \$
  },
  sensitive=false, % keywords are not case-sensitive
  morecomment=[l][\color{codegreen}]{-},
  morestring=[b]",
}

\lstdefinelanguage{form}
{
  morekeywords={
    id,
    if,
    else,
    elsif,
    endif,
    occurs,
    exit,
    repeat,
    endrepeat,
  },
    otherkeywords={
     \#define
    },
  sensitive=false, % keywords are not case-sensitive
  morecomment=[f][\color{gray}][0]{*},
  morecomment=[f][\color{magenta}][0]{*--\#[},
  morecomment=[f][\color{magenta}][0]{*--\#]},
  morecomment=[f][\color{blue}][0]{.},
  morestring=[b]",
}

\lstdefinestyle{code}{
    backgroundcolor=\color{backcolour},   
    commentstyle=\color{codegray},
    keywordstyle=\color{magenta},
    stringstyle=\color{codepurple},
    basicstyle=\ttfamily\footnotesize,
    breakatwhitespace=false,         
    breaklines=true,                 
    captionpos=b,                    
    keepspaces=true,                 
    numbersep=5pt,                  
    showspaces=false,                
    showstringspaces=false,
    showtabs=false,                  
    tabsize=2,
    title=\lstname,
    belowskip=-1em,
    aboveskip = 1em
}

\begin{document}
\title{Three-loop topology analysis of neutral B-meson mixing with \texttt{tapir}}

\author{Marvin Gerlach}

\address{Institut f\"ur Theoretische Teilchenphysik, Karlsruhe Institute of Technology (KIT), Wolfgang-Gaede-Straße 1, 76128 Karlsruhe, Germany}

\ead{gerlach.marvin@protonmail.com}

\begin{abstract}
Modern advances in particle physics depend strongly on the usage of reliable computer programs. In this context two issues become important: The usage of powerful algorithms to handle the amount of evaluated data properly, and a software architecture capable to overcome the problems of maintainability and extendability.
We present our approach to such a computer program, called \texttt{tapir}. This tool assists computations in perturbative quantum field theory in many ways. Such calculations often involve the evaluation of a large amount of Feynman diagrams with multiple loops. \texttt{tapir} helps in reducing the number of diagrams, and the resulting integrals thereof, by identifying and minimizing their topological structure. 
We will focus on a three-loop calculation which is needed for the next-to-next-to leading order predictions of neutral $B$-meson systems.  We show how \texttt{tapir} can be utilized for this kind of calculation.
\end{abstract}

\section{Introduction}

The outstanding experimental effort of recent years led to an unreached precision in flavor observables. Among the most prominent ones is the mixing of neutral meson states such as the $B_s$ and $B_d$. We take in this work only $B_s$ into account, the extension to $B_d$ is straightforward. A more elaborate overview for this topic is given in e.g. Ref.~\cite{buras_2020}.

$B_s$ can oscillate in its antiparticle state $\overline{B}_s$ via box diagrams. Therefore, the mass eigenstates differ from the flavor eigenstates. A few observables for this flavor oscillation process are of particular interest in connection to precision tests of the Standard Model. They can be derived from the transition amplitude from one flavor eigenstate to another, which we denote as
\begin{equation}
    \begin{aligned}
        -i (2\pi)^4\delta^{(4)}\left(p_{\text{i}}-p_{\text{j}}\right)\Sigma_{\text{ij}}^s ~=~ \frac{\bra{\text{i}} i \mathcal{T} \ket{\text{j}}}{2 M_{B_s}}\,.
    \end{aligned}
\end{equation}
i and j describe either a particle state $B_s$ or an antiparticle state $\overline{B}_s$. $M_{B_s}$ is the average mass of the meson.

Especially the off-diagonal elements of this amplitude are interesting for the mixing due to the connection to the mass and decay width matrix:
\begin{equation}
    \begin{aligned}
        M_{12}^s ~&=~ \frac{\Sigma_{12}^s + (\Sigma_{21}^{s})^*}{2} ~\equiv~ \text{Disp}\!\left(\Sigma_{12}^s\right)\,,\\
        \Gamma_{12}^s ~&=~ i\left(\Sigma_{12}^s - (\Sigma_{21}^{s})^*\right) ~\equiv~ 2 \text{Abs}\!\left(\Sigma_{12}^s\right)\,.
    \end{aligned}
\end{equation}

In order to compute these off-diagonal matrix elements, it is hence necessary to evaluate the \textit{dispersive} and the \textit{absorptive} part of the process $\overline{B}_s \rightarrow B_s$. They correspond to the real and imaginary part of the scattering amplitude where the CKM matrix elements are factored out. According to the optical theorem, only virtual particles contribute to $\Gamma_{12}^s$ which can be produced on-shell. For $M_{12}^s$ exists no such restriction. Hence, $\Gamma_{12}^s$ is straightforward to compute when connected to the calculation of the discontinuity of the complex plane, and with it to Cutkosky's \textit{cutting rules}~\cite{1960JMP.....1..429C}.

From $M_{12}^s$ and $\Gamma_{12}^s$ we are able to compute three physical observables: The mass difference between heavy (H) and light (L) mass eigenstates,
\begin{equation}
    \begin{aligned}
        \Delta M_s ~=~ M_{\text{H}}^s - M_{\text{L}}^s ~=~ 2 |M_{12}^s|\,,
    \end{aligned}
\end{equation}
the decay width difference,
\begin{equation}
    \begin{aligned}
        \Delta \Gamma_s ~=~ \Gamma_{\text{L}}^s - \Gamma_{\text{H}}^s ~\approx~ 2 |\Gamma_{12}^s|\,,
    \end{aligned}
\end{equation}
and the CP asymmetry
\begin{equation}
    \begin{aligned}
        a_{\text{fs}}^s ~=~ \text{Im}\left(\frac{\Gamma_{12}^s}{M_{12}^s}\right)\,.
    \end{aligned}
\end{equation}
$a_{\text{fs}}^s$ is a measure of CP breaking in the mixing by quantifying how much the mass eigenstates differ from the CP eigenstates. 

$\Delta \Gamma_s$ is computed in an effective field theory with the effective interaction Hamiltonian $\mathcal{H}^{\Delta B = 1}$:
\begin{equation}
    \label{eq:gamma12}
    \begin{aligned}
        \Gamma_{12}^s ~=~ \frac{1}{2 M_{B_s}}\text{Abs}\!\left(i \bra{B_s} \int\!\text{d}^4x \ \mathcal{H}^{\Delta B = 1}(x) \mathcal{H}^{\Delta B = 1}(0) \ket{\overline{B}_s}\right)\,.
    \end{aligned}
\end{equation}

For $\mathcal{H}^{\Delta B = 1}$ we choose the basis of Ref.~\cite{Chetyrkin:1997gb} which is closed under renormalization when including QCD effects. It also has the advantage of preventing  traces including $\gamma_5$. The numerically most relevant effective operators are the so-called \textit{current-current} operators. Taking only those into account, we have
\begin{equation}
    \begin{aligned}
        \mathcal{H}^{\Delta B = 1} ~=~ \frac{4 G_F}{\sqrt{2}} \sum_{q_1,q_2}  V_{q_1b} V_{q_2s}^* \left( C^{q_1 q_2}_1 P^{q_1 q_2}_1 + C^{q_1 q_2}_2 P^{q_1 q_2}_2 \right)\,.
    \end{aligned}
\end{equation}
The sum iterates over $q_1,q_2 \in \{u,c\}$. The operators are given by
\begin{equation}
    \label{eq:cc-operators}
    \begin{aligned}
        P^{q_1 q_2}_1 ~&=~ \left(\overline{s} \ T^a \gamma_{\mu} P_L \ q_2\right) \left(\overline{q_1} \ T^a \gamma^{\mu} P_L \ b\right)\,,\\
        P^{q_1 q_2}_2 ~&=~ \left(\overline{s} \ \gamma_{\mu} P_L \ q_2\right) \left(\overline{q_1} \ \gamma^{\mu} P_L \ b\right)\,,
    \end{aligned}
\end{equation}
with $P_L = (1-\gamma_5)/2$.

\Cref{eq:gamma12} can be even further simplified by using the \textit{heavy quark expansion} (HQE). This is an operator product expansion around the small parameter $\Lambda/m_b$. It translates the non-local hadronic matrix element of \cref{eq:gamma12} to local operator insertions with new Wilson coefficients $H$ and $\tilde{H}_S$. We hence get
\begin{equation}
    \begin{aligned}
        \Gamma_{12}^s ~=~ \frac{G_F^2 m_b^2}{24 \pi M_{B_s}} \left( H \bra{B_s} Q \ket{\overline{B}_s} + \tilde{H}_S \bra{B_s} \tilde{Q}_S\ket{\overline{B}_s} \right)\,,
    \end{aligned}
\end{equation}
with
\begin{equation}
    \begin{aligned}
        Q ~=~ 4 \left(\overline{s}_{\text{i}} \ \gamma_{\mu} P_L \ b_{\text{i}}\right) \left(\overline{s}_{\text{j}} \ \gamma^{\mu} P_L \ b_{\text{j}}\right)\,, 
        \quad
        \tilde{Q}_S ~=~ 4 \left(\overline{s}_{\text{i}} \ P_L \ b_{\text{j}}\right) \left(\overline{s}_{\text{j}} \ P_L \ b_{\text{i}}\right)\,.
    \end{aligned}
\end{equation}

\begin{figure}
    \begin{center}
        \includegraphics[width=\textwidth]{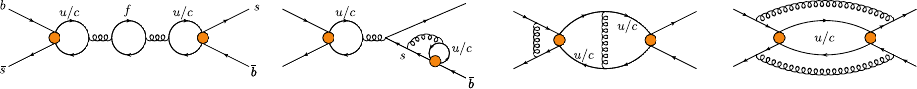}
    \end{center}
    \caption{Sample diagrams which contribute at NNLO to $\Delta \Gamma_s$ within the effective $\Delta B = 1$ theory.}
    \label{fig:bmix-dias}
\end{figure}

Corrections to \cref{eq:gamma12} including only current-current operators are currently known up to $\mathcal{O}(\alpha_s)$~\cite{Beneke:1998sy,Ciuchini:2003ww,Beneke:2003az,Lenz:2006hd} and only partially known to $\mathcal{O}(\alpha_s^2)$~\cite{Asatrian:2017qaz,Asatrian:2020zxa}. To complete the missing NNLO contributions, diagrams as in \cref{fig:bmix-dias} have to be considered.

\section{Diagram and topology analysis}
Analyzing the diagrams of \cref{fig:bmix-dias} gives a hint which problems may occur. In order to allow for one-particle reducible diagrams (diagram~1), we have to use the options ``\texttt{notadpoles}'' and ``\texttt{onshell}'' of \texttt{qgraf}~\cite{Nogueira:1991ex} during diagram generation. Unfortunately, ``\texttt{onshell}'' also discards diagram~2. Therefore, we define a second diagram class where only the option ``\texttt{offshell}'' is specified. We then use a special diagram filter of \texttt{tapir}~\cite{Gerlach:2022qnc} to remove the unwanted diagrams. We do this by specifying the following in the \file{config} file:
\begin{lstlisting}[language=tapirconf]
* tapir.filter external_self_energy_bridge : false
\end{lstlisting}
This filter excludes all diagrams whose external legs have self-energy corrections. Flavor mixing of the self-energy, as in diagram~2, is explicitly not effected.

\texttt{qgraf} generates $\mathcal{O}(4500)$ three-loop diagrams for the first diagram class. Unfortunately, \texttt{qgraf} can only distinguish between interactions with different particles, i.e.~$P_1^{cc}$ and $P_2^{cc}$ cannot be differentiated. The separation is first applied if symbolic graph representations (\file{dia}/\file{edia}) are generated. Then, \texttt{tapir} iterates over all possible interaction combinations as stated in the \file{vrtx} file. Thus, the number of diagrams increases to $\mathcal{O}(17000)$ if all current-current operators of \cref{eq:cc-operators} are regarded. The second class starts with $\mathcal{O}(3000)$ diagrams generated by \texttt{qgraf}, but reduces to $\mathcal{O}(2000)$ when applying the described diagram filter. The actual number of remaining diagrams of the second class with distinguished operators is $\mathcal{O}(8000)$.

A further simplification concerns the specific kinematics of the process $\overline{b}(q_1) s(q_2) \rightarrow \overline{s}(q_3) b(q_4)$. Since we match the results of two effective field theories to get $H$ and $\tilde{H}_S$, the result must not depend on the kinematics of the external particles. Thus, we can choose the three-momenta of the $s$-quarks as $\vec{q}_2 = \vec{q}_3 = 0$. In the limit $m_s \rightarrow 0$, this kinematic is topologically equal to diagrams where the external $s$-lines are removed. We can make use of this in \texttt{tapir} with
\begin{lstlisting}[language=tapirconf]
* tapir.external_momentum q2:0
* tapir.external_momentum q3:0
\end{lstlisting}
The external momenta are labeled according to the order of how the external particles were declared in \file{qgraf.dat}. Additionally, we specify that the $b$- and $c$-quarks are massive:
\begin{lstlisting}[language=tapirconf]
* tapir.mass fb:M1
* tapir.mass fc:M2
\end{lstlisting}

After specifying a \file{prop} and \file{vrtx} file with all symbolic Feynman rules, we analyze and minimize the occurring topologies. We call \texttt{tapir} with the following command line options:
\begin{lstlisting}[language=shell]
$ /path/to/tapir -c class1.conf -q qlist.3 -m -t class1.top -k 8
\end{lstlisting}
The ``\texttt{-c}'' argument specifies the \file{config} file. The argument of ``\texttt{-q}'' defines the \texttt{qgraf} output file that was generated using the style file which comes with \texttt{tapir}. ``\texttt{-m}'' minimizes the topologies of the diagrams and tries to map them onto each other. The option ``\texttt{-t}'' outputs the remaining topologies as a \file{topsel} file. This file format was adopted from \texttt{q2e} and \texttt{exp} ~\cite{Harlander:1998cmq,Seidensticker:1999bb,q2eexp} and serves as the standard format for graph topologies around \texttt{tapir}. The last command line option ``\texttt{-k}'' demands \texttt{tapir} to use 8 cores for parallel evaluation. The execution per diagram class takes $\mathcal{O}(\SI{10}{s})$ on a modern desktop PC.

\section{Topology minimization}
The implemented minimization starts with extracting the topological features of each diagram. This includes the connections of the underlying graph, i.e.~which vertices are connected by which edge. It also takes the \textit{edge coloring} into account, i.e.~the mass of the propagating particle. This information is then encoded in a label which is unique for a given Feynman graph topology: the \textit{Nickel index}~\cite{Nickel:1977}. With this index at hand it is straightforward to identify a priori different graphs, by using it as a hash function and comparing the generated hashes. This procedure enables a minimization in a highly parallelized manner. 

The two diagram classes defined above lead minimized to $\mathcal{O}(1000)$ and $\mathcal{O}(400)$ remaining topologies, respectively. In the next step we want to combine both topology sets, and reduce them even further. For this purpose, we create a new \file{config} file with the following entries:
\begin{lstlisting}[language=tapirconf]
* tapir.topo_ignore_bridges
* tapir.topo_remove_duplicate_lines
\end{lstlisting}

Here, we use a set of destructive topology filters, i.e.~they change the graph topology, but do not affect the underlying integral families.
The option ``\texttt{tapir.topo\_ignore\_bridges}'' removes lines from a topology which do not carry a loop momentum, so-called \textit{bridges}. ``\texttt{tapir.topo\_remove\_duplicate\_lines}'' removes lines which have the exact same momentum and mass as another line.

As a next step, we want to use the topological information to build symbolic integral families. For this, we generate a \texttt{FORM}~\cite{Ruijl:2017dtg} file that combines the scalar loop-dependent factors to a topology function. This \file{topology} file can be embedded in a \texttt{FORM}-based setup, as it is common for multi-loop calculations. Before including this file, some kind of tensor reduction has to be applied to the symbolic expressions of the diagrams generated by \texttt{tapir}. The generation of the \file{topology} file can be varied by several options. For our case, we may use
\begin{lstlisting}[language=tapirconf]
* tapir.topo_complete_momentum_products
\end{lstlisting}
to replace numerator functions of the integral family by a standard propagator function. For example, it may happen that the product $q.k$ cannot be expressed through the remaining propagators. Thus, $q.k$ would be part of the integral family. The presented option replaces the product with
\begin{equation}
    q.k ~=~ \frac{1}{2} \left((q + k)^2 - q^2 - k^2\right)\,.
\end{equation}
The combination $q + k$ is then replaced by an ancillary momentum $P_1$, and $1/(-P_1^2)^n$ is included in the integral family. The minus sign is used for convenience.

We finally can call tapir once again with
\begin{lstlisting}[language=shell]
$ /path/to/tapir -c combine.conf -i class12.top -m -t final.top -f include/ -k 8 -pf
\end{lstlisting}

This time, we read a concatenation of the previously created \file{topsel} files. One has to make sure that the topologies of the different classes can be distinguished using the option ``\texttt{tapir.topology\_name}'' beforehand. The option ``\texttt{-f}'' creates a folder with the \file{topology} files in addition to a \file{topology list} file containing all integral family definitions in a \texttt{Mathematica} readable format. The latter can be used as input for reduction programs like \texttt{FIRE}~\cite{Smirnov:2019qkx}.

Additionally, we use the option ``\texttt{-pf}'' which usually performs a partial fraction decomposition of the final integral families. Thus, we ensure to get only integral families with linear independent denominator functions. This is a necessary pre-condition for integration-by-parts algorithms. The combination of both diagram classes with the mentioned topology filters results in $\mathcal{O}(900)$ topologies. It turns out, the partial fraction decomposition does not apply since the options \texttt{topo\_ignore\_bridges} and \texttt{topo\_remove\_duplicate\_lines} already remove all occurring linear dependencies in the integral families. Nevertheless, using ``\texttt{-pf}'' is always a good precaution.

The remaining $\mathcal{O}(900)$ topologies are the result of only a coarse minimization. A program like \texttt{exp} can map the diagrams on even less topologies. Thus, our naive minimization procedure can be regarded as a relatively fast ``pre-filtering''.

The generated \file{topology} files, e.g. for the topology of diagram 4 in \cref{fig:bmix-dias}, have the following form:

\begin{lstlisting}[language=form]
* Reducible numerator momentum replacements
id p3 = -p4 - q1;
id p1 = p4 - p5 + p6;
id p2 = p4 - p5 + p6 + q1;
id P1 = p5 + p6;
id P2 = p4 + p6;
id P3 = p5 + q1;
.sort

* Numerator momentum product replacements
id p6.q1 = -p1.p1/2 + p2.p2/2 - p3.p3/2 + p4.p4/2 + p5.q1;
id p4.q1 = p3.p3/2 - p4.p4/2 - q1.q1/2;
id p4.p5 = -p1.p1/2 + p4.p4/2 + p4.p6 + p5.p5/2 - p5.p6 + p6.p6/2;
id p5.p6 = (P1.P1 - p5.p5 - p6.p6)/2;
id p4.p6 = (P2.P2 - p4.p4 - p6.p6)/2;
id p5.q1 = (P3.P3 - p5.p5 - q1.q1)/2;
.sort

* Define massive propagators
id p2.p2 = -1/s2m1 + M1^2;
id p4.p4 = -1/s4m1 + M1^2;
id p5.p5 = -1/s5m2 + M2^2;
id p6.p6 = -1/s6m2 + M2^2;
.sort

* Combine to scalar topology function
id s6m2^n0? * s5m2^n1? * s4m1^n2? * s2m1^n3? * 1/p1.p1^n4? * 1/p3.p3^n5? * 1/P1.P1^n6? * 1/P2.P2^n7? * 1/P3.P3^n8? = 
    (-1)^n4 * (-1)^n5 * (-1)^n6 * (-1)^n7 * (-1)^n8 * 
    DB1TopoNNLOdr3l4380(n0,n1,n2,n3,n4,n5,n6,n7,n8);
.sort
\end{lstlisting}

Herewith a tensor reduced, symbolic diagram is straightforwardly replaced by an integral family in \texttt{FORM}. The corresponding \file{topology list} entry is given by

\begin{lstlisting}[language=C++]
{"DB1TopoNNLOdr3l4380", {M2^2 - p6^2, M2^2 - p5^2, M1^2 - p4^2, M1^2 - (p4 - p5 + p6 + q1)^2, -(p4 - p5 + p6)^2, -(p4 + q1)^2, -(p5 + p6)^2, -(p4 + p6)^2, -(p5 + q1)^2}, {p4, p5, p6}}
\end{lstlisting}
which is in the form of 
\begin{lstlisting}[language=C++]
{"Family name", List of denominators, List of loop momenta}
\end{lstlisting}

\section{Conclusion}
We have shown how the program \texttt{tapir} can be used to simplify Feynman graph topologies and how they can be expressed as symbolic integral families. We have considered the example of $B_s-\overline{B}_s$ mixing at three-loop order. The program was already used in similar two-loop calculations~\cite{Gerlach:2021xtb,GNSS:penguin}.

The program and an extensive documentation are freely available at the \texttt{gitlab} page: \url{https://gitlab.com/F.Herren/tapir}.

\ack
I thank Matthias Steinhauser for careful proofreading of this article. I also want to thank Florian Herren and Martin Lang for the good collaborative work on \texttt{tapir}.

This research was supported by the Deutsche Forschungsgemeinschaft (DFG, German Research Foundation) under grant 396021762 — TRR 257 ``Particle Physics Phenomenology
after the Higgs Discovery''.

\section*{References}
\bibliography{references}

\end{document}